\documentclass[twocolumn,showpacs,preprintnumbers,amsmath,amssymb,aps,prl,superscriptaddress]{revtex4-1}

\usepackage{color}
\usepackage{graphicx}
\usepackage{textcomp} 
\usepackage{subeqnarray}
\usepackage[caption=false]{subfig}

\newcommand \be  {\begin{equation}}
\newcommand \beno  {\begin{equation*}}
\newcommand \bea {\begin{eqnarray} \nonumber }
\newcommand \ee  {\end{equation}}
\newcommand \eeno  {\end{equation*}}
\newcommand \eea {\end{eqnarray}}
\begin{document}
\title{Crossover from Linear to Square-Root Market Impact}
\author{Fr\'ed\'eric Bucci}
\affiliation{Scuola Normale Superiore di Pisa, Piazza dei Cavalieri 7, 56126 Pisa, Italy}
\author{Michael Benzaquen}
\affiliation{Ladhyx UMR CNRS 7646 \&  Department of Economics, Ecole Polytechnique, 91128 Palaiseau Cedex, France}
\affiliation{Capital Fund Management, 23-25, Rue de l'Universit\'e 75007 Paris, France}
\author{Fabrizio Lillo}
\affiliation{Department of Mathematics, University of Bologna, Piazza di Porta San Donato 5, 40126 Bologna, Italy}
\author{Jean-Philippe Bouchaud}
\affiliation{Capital Fund Management, 23-25, Rue de l'Universit\'e 75007 Paris, France}
\affiliation{CFM-Imperial Institute of Quantitative Finance, Department of Mathematics, Imperial College, 180 Queen's Gate, London SW7 2RH}

\date{\today}

\begin{abstract}
Using a large database of 8 million institutional trades executed in the U.S. equity market, we establish a clear crossover between a linear market impact regime and a square-root regime as a function of the volume of the order. Our empirical results are remarkably well explained by a recently proposed dynamical theory of liquidity that makes specific predictions about the scaling function describing this crossover. Allowing at least two characteristic time scales for the liquidity (``fast'' and ``slow'') enables one to reach quantitative agreement with the data. 
\end{abstract}
\pacs{}
\maketitle

% insert suggested PACS numbers in braces on next line
%\pacs{}
% insert suggested keywords - APS authors don't need to do this
%\keywords{}

%\maketitle must follow title, authors, abstract, \pacs, and \keywords
% body of paper here - Use proper section commands
% References should be done using the \cite, \ref, and \label commands
\section{}

Financial markets sputter enormous amounts of data that can now be used to test scientific theories at levels of precision comparable to those achieved in physical sciences (see, e.g. \cite{Takayazu} for a recent example). Among the most remarkable empirical findings in the last decades is the ``square-root impact law'', which quantifies how much prices are affected, {\it on average}, by large buy or sell orders, usually executed as a succession of smaller trades. Such a succession of small trades, all executed in the same direction (either buys or sells) and originating from the same market participant, is called a \emph{metaorder}. A metaorder of total size $Q$ impacts the price as $\sim \sqrt{Q}$ and not proportionally to $Q$ as naively expected and actually predicted by classical economics arguments \cite{Kyle}. The square-root law is surprisingly universal: it is found to be to a large degree independent of details such as the asset class, time period, execution style and market venues \cite{Torre, Almgren, Engle, Moro, Toth, Brokmann, Zarinelli, Bacry, Bonart, Toth2, Bucci, Frazzini}. In particular, the advent of electronic markets and High Frequency Trading has {\it not} altered the square-root behaviour, in spite of radical changes in the microstructure of markets. 

The universality of this square-root law, together with its insensitivity to the high frequency dynamics of prices, suggests that its interpretation should lie in some general properties of the low frequency, large scale dynamics of liquidity \cite{TQP}. In fact, the publicly displayed liquidity at any given time is usually very small -- typically on the order of $10^{-2}$ of the total daily transaction volume in stock markets. Financial markets are the arena of a collective hide-and-seek game between buyers and sellers, resulting in a somewhat paradoxical situation where the total quantity that markets participants intend to trade is very large ($0.5 \%$ of the total market capitalisation changes hands every day in stock markets) while most of this liquidity remains hidden, or ``latent''. These observations have lead to the development of a physics inspired, ``locally linear order book'' (LLOB) model for the coarse-grained dynamics of latent liquidity \cite{Toth,MTB,Donier,TQP}, which naturally explains why the impact of metaorders grows like the square-root of its size in a certain regime of parameters \cite{Donier}. But this LLOB model also suggests that for a given execution time $T$, the very small $Q$ regime should revert to a linear behaviour. The model in fact predicts the detailed shape of the crossover between linear and square-root impact. Deviations from a pure square-root were observed in \cite{Zarinelli}, where the authors fitted the data with a logarithmic function $\ln(a + bQ)$, which indeed behaves linearly for small arguments. 

The aim of the present letter is to test for the first time the detailed theoretical predictions {of a crossover from linear to square root impact }using the very large ANcerno \footnote{\uppercase{A}Ncerno Ltd (formerly the Abel Noser Corporation) is a widely recognized consulting firm that works with institutional investors to monitor their equity trading costs. Its clients include many pension funds and assets managers. Previous academic studies that use ANcerno data to investigate the market impact at different times scales includes \cite{Zarinelli, Bucci}. See {\tt www.ancerno.com} for details.} database of metaorders, executed on the US equity market and issued by a diversified set of institutional investors. We find that the crossover between linear and square-root impact is well described by the theory, {albeit the transaction volume at the crossover point is much smaller than the one predicted by the theory.}

We argue that this can be accounted for by the coexistence of ``slow'' and ``fast'' agents in financial markets. Fast agents contribute to the total transaction volume but are unable to offer resistance against the execution of large metaorders. Therefore, only slow agents are able to dampen market impact and only their contribution is relevant for shaping up the square-root law. We recall how the LLOB model can be augmented to account for multiple agent frequencies \cite{Benzaquen}, and compute the impact crossover function within this extended framework, resulting in a remarkably good fit of the data.

Let us first briefly recall the basic ingredients of the LLOB model, as well as its main predictions. The fundamental quantity of interest is the density $\varphi(x,t)$ of latent orders around price $x$ at time $t$. Conventionally, one can choose $\varphi$ to be positive for buy latent orders (corresponding to $x < p(t)$, where $p(t)$ is the current transaction price) and negative for sell latent orders (corresponding to $x > p(t)$). As argued in \cite{Toth,MTB,Donier}, the coarse-grained dynamics of the latent liquidity close to the current price is well described by the following equation:
\begin{equation}
\partial_t \varphi= D\partial_{xx} \varphi - \nu \varphi + \lambda~\textrm{sign}(y)+m~\delta(y) \ ,
\label{eqB}
\end{equation}
where $y:=p(t)-x$, and $\nu$ describes order cancellation, $\lambda$ new order deposition and $D\partial_{xx}$ limit price reassessments. The final ``source'' term corresponds to a metaorder of size $Q$ executed at a constant rate $m=Q/T$, corresponding to a flux of orders localized at the transaction price $p(t)$. In the absence of a metaorder ($m=0$), Eq. (\ref{eqB}) admits a stationary solution in the price reference frame, which is {\it linear} when $y$ is small (hence the name ``LLOB''):
\be
\varphi_{\text st}(y) = {\cal L} y\ ,
\ee
where ${\cal L}=\lambda/\sqrt{D \nu}$ is a measure of liquidity. The linear behaviour of the latent liquidity close to the transaction price is in fact a generic result, that holds much beyond the simple setting defined by Eq. (\ref{eqB}), while being at the origin of the square-root impact law \cite{Toth,MTB,Donier,TQP}. The total transaction rate $J$ is simply given by the flux of orders through the origin, i.e. $J := D \partial_y \varphi_{\text st}|_{y=0} = D {\cal L}$. 

In the limit of a slow latent order book (i.e. $\nu T \ll 1$), the price trajectory $p_m(t)$ during the execution of the metaorder (obtained as the solution of $\varphi(p_m,t)=0$) is given by the following self-consistent expression \cite{Donier}:
\begin{align}
%\begin{split}
\hspace{2.5cm} p_m(t) =& p_0(t) + y(t), \\  & \hspace{-4cm}y(t) = \frac{m}{{\cal L}} \int_0^t \frac{{\rm d}s}{\sqrt{4 \pi {D} (t-s)}} \exp\left[{-\frac{(y(t) - y(s))^2}{4{D}(t-s)}}\right],
%\end{split}
\end{align}
where $p_0(t)$ is the price trajectory in the absence of the metaorder that starts at $t=0$ and ends at $t=T$. Price impact is then defined as $I:=y(T)$, and is found to be given by:
\be\label{eq:impact}
{I}(Q) = \sqrt{\frac{DQ}{J}}\, {\cal F}(\eta) \ ,\quad\mbox{with} \quad \eta:=\frac{Q}{JT}\ ,
\ee
where $\eta$ is the {\it participation rate} and the scaling function ${\cal F}(\eta) \approx \sqrt{\eta/\pi}$ for $\eta \ll 1$ and $\approx \sqrt{2}$ for $\eta \gg 1$. Hence, $I(Q)$ is linear in $Q$ for small $Q$ at fixed $T$, and crosses over to a square-root for large $Q$. Note that in the square-root regime, impact is predicted to be {\it independent} of the execution time $T$. Many other results, such as the decay of impact for $t > T$, have been derived and discussed in \cite{Donier,Benzaquen}. 

We now turn to the ANcerno database to see how well Eq. (\ref{eq:impact}) is supported empirically. Our sample covers for a total of $880$ trading days, from January 2007 to June 2010 and we follow the cleaning procedure introduced in \cite{Zarinelli} to remove possible spurious effects. The sample is represented by around 8 million metaorders uniformly distributed in time and market capitalization \footnote{The sample represents around the 5\% of the total market volume}. Each metaorder in the database is characterised by a broker label, a stock symbol, the total number of traded shares $Q$, the sign $\epsilon =\pm 1$ (buy/sell), the start-time $t_\mathrm{s}$ and the end-time $t_e$ of its execution. In line with the definition given above, and following \cite{Zarinelli}, the participation rate is given by $\eta=Q/V_T$ where $V_T=V(t_e)-V(t_\mathrm{s})$ is the total volume traded in the market during the metaorder execution. In order to compare different stocks with very different daily volumes, we shall measure $Q$ in units of the corresponding daily volume $V_{\text{d}}$, and introduce the volume fraction $\phi:=Q/V_{\text{d}}${, which in the model notation is equal to $Q/JT_{\text{d}}$, where $T_{\text{d}}=1$ day}. We will also measure execution in relative volume time and redefine the execution time $T$ as $T:=(V(t_e)-V(t_\mathrm{s}))/V_{\text{d}}$. Finally, we introduce rescaled log-prices as $p(t):=(\log~P(t))/\sigma_{\text{d}}$, where $\sigma_{\text{d}}=(P_{\mathrm{high}}-P_{\mathrm{low}})/P_{\mathrm{open}}$ is the daily volatility estimated from the daily high, low and open prices $P(t)$. 

The average price impact $I(Q)$ for a given executed volume $Q$, as studied in most previous studies, is defined as:
\begin{equation}
I(Q)=\mathbb{E}[\epsilon \cdot (p_{e}-p_{s}) \vert Q]\ ,
\label{Eq:imp}
\end{equation}
where $p_\mathrm{s}$, $p_e$ are, respectively, the mid-price at the start and at the end of the metaorder. As shown in \cite{Zarinelli, Bucci}, the ANcerno data
confirms that $I(Q)$ is close to a square-root in an intermediate regime of volume fraction $10^{-3} \lesssim \phi \lesssim 10^{-1}$, but shows an approximate linear behaviour for 
smaller volume fractions $\phi \lesssim 10^{-3}$. Note that data for large volume fractions $\phi \gtrsim 10^{-1}$ are difficult to interpret, as they are prone to strong conditioning effects. 

In order to test directly Eq. (\ref{eq:impact}), we  estimate the scaling function ${\cal F}(\eta)$ by dividing the data into evenly populated bins of constant participation rate $\eta$ and compute the conditional expectation of $\epsilon(p_{e}-p_{s})/\sqrt{\phi}$ for each bin. {According to the LLOB model this expectation is equal to $\sqrt{D/\sigma^2_d} {\cal F}(\eta) $}. Here and in the following, error bars are determined as standard errors. 

\begin{figure}[ht!]
\centering
\includegraphics[width=0.8\linewidth]{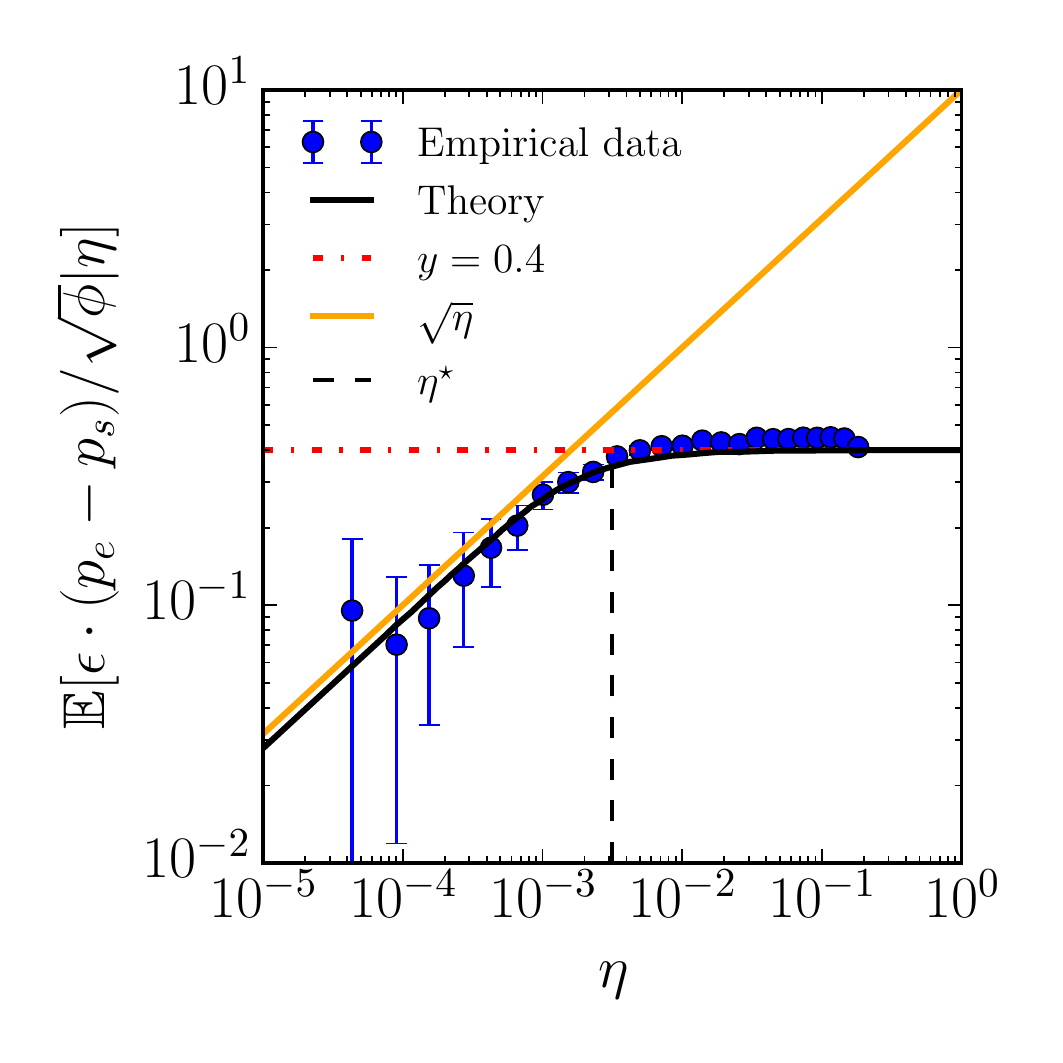}
\caption{Empirically determined scaling function ${\cal F}(\eta)$ vs. participation rate $\eta$. The data (blue points) interpolates between a $\sqrt{\eta}$ behaviour observed at small participation rates and an asymptotically constant regime $\approx 0.4$ for large $\eta$, i.e. for $\eta \gtrsim \eta^\star$ with $\eta^{\star} \approx 3.15\times10^{-3}$. Black line: prediction of LLOB, with an adjusted crossover $\eta^\star:=J_\mathrm{s}/J_\mathrm{f}$ allowing for the existence of two categories of agents (``fast'' and ``slow''). The data points are obtained by restricting to metaorders with sufficient large order size, i.e. $\phi \gtrsim 10^{-5}$.}
\label{figDonier}
\end{figure}

The results are shown in Fig. \ref{figDonier} and are, up to a rescaling of both the $x$- and $y$- axis, remarkably well accounted for by the LLOB function ${\cal F}(\eta)$, that describes the crossover between a linear-in-$Q$ regime for small participation rates $\eta$, and a $T$-independent, $\sqrt{Q}$ regime at large $\eta$. Whereas a linear regime for small $Q$'s was already reported in \cite{Zarinelli,Bucci}, the scaling analysis provided here has not been attempted before. The fact that impact in $\sqrt{Q}$ regime chiefly depends on $Q$ but not on $T$ is compatible with the results of \cite{CFMunpub,Zarinelli}, but contradicts theories that assigns the $\sqrt{Q}$ dependence to {\it duration} of the metaorder, as in Refs. \cite{Torre,Zhang,Gabaix}.\footnote{Directly regressing the impact as $\sqrt{\phi} T^{-\beta}$ in the $\eta > \eta^\star$ regime yields $\beta = - 0.04 \pm 0.02$, confirming the near independence of the impact on the execution time $T$; {while in the $\eta < \eta^\star$ regime the regression as $\phi T^{-\beta}$ yields $\beta = 0.45 \pm 0.05$, in close agreement with the LLOB prediction $\beta=1/2$.}}

However, whereas the crossover between the two regimes should occur around $\eta^\star=1$ within the original LLOB model, empirical data points towards a much smaller value $\eta^\star \sim 10^{-3}$. This is actually consistent with the fact that all the empirical evidence for the square-root law reported in the literature concern moderate participation rates (typically in the range $10^{-3} - 10^{-1}$, see e.g. \cite{Almgren,Brokmann,Bucci,Frazzini}) but never in a regime where the volume of the metaorder becomes larger than the rest of the market, as would be requested within the LLOB specification. Note that in our sample, $70 \%$ of the metaorders are such that $\eta >\eta^\star$.

In order to account for this large discrepancy in the value of $\eta^\star$, we shall consider the extended the LLOB model recently proposed by two of us to include agents with different time horizons \cite{Benzaquen}. In the simplest case of a bi-modal distribution of agents (``fast'' and ``slow''), the LLOB formalism can be generalized to describe two latent order book densities, $\varphi_\mathrm{s}(x,t)$ for the slow liquidity and $\varphi_\mathrm{f}(x,t)$ for the fast liquidity -- for example provided by High Frequency Traders. The corresponding dynamical equations read \cite{Benzaquen}: 
\begin{equation}
\partial_t \varphi_{\circ}= D_{\circ}\partial_{xx} \varphi_{\circ} - \nu_{\circ} \varphi_{\circ} + \lambda_{\circ}\textrm{sign}(y)+m_{\circ}(t)\delta(y) \ ,
\label{eqS}
\end{equation}
where $y = p(t)-x$ and $\circ = \text{s, f}$, and where $m_\mathrm{s}(t)$ (resp. $m_\mathrm{f}(t)$) is the fraction of the metaorder absorbed by the slow (resp. fast) traders, with $m_{s}(t)+m_{f}(t)=m$. We allow the activity rate
of the two categories of agents to be different through the coefficients $D_{\circ}$, $\nu_{\circ}$ and $\lambda_{\circ}$. The interesting limit for our purpose is:
\begin{itemize}
\item $J_\mathrm{s} \ll J_\mathrm{f}$ and $m \ll J_\mathrm{f}$, where $J_{\circ} = \lambda_{\circ} \sqrt{D_{\circ}/\nu_{\circ}}$. These inequalities mean that (i) the {total transaction rate} $J=J_\mathrm{s}+J_\mathrm{f}\approx J_\mathrm{f}$ is dominated by fast traders and (ii) the flux corresponding to the metaorder is small compared to the {total transaction rate} of the market, as with most metaorders executed in liquid markets. 
\item $\nu_\mathrm{s} T \ll 1$ and $\nu_\mathrm{f} T \gg 1$. {As shown in \cite{Benzaquen} this implies that slow, persistent agents are able to resist to the impact of the metaorder, whereas fast agents are playing the role of intermediaries, only lubricating the high-frequency activity of markets.}
\end{itemize}
This double-frequency model can be solved exactly in some limits \cite{Benzaquen}. One should distinguish two cases, depending on whether the execution time $T$ is larger or smaller
than a certain $T^\dagger := \nu_\mathrm{f}^{-1} \eta^{\star -2} D_\mathrm{s}/D_\mathrm{f}$, where $\eta^{\star}:=J_\mathrm{s}/J_\mathrm{f}$. For $T > T^\dagger$, the scaling result Eq. (\ref{eq:impact}) is simply modified as:  
\be\label{eq:impact2}
{I}(Q) = \sqrt{\frac{D_\mathrm{s}Q}{J_\mathrm{s}}}\, {\cal F}\left(\frac{\eta}{\eta^\star}\right).
\ee
For $T < T^\dagger$, this result is further multiplied by $\sqrt{T/T^\dagger}$, with a shifted crossover point $\eta^\star \to \eta^\star T^\dagger/T$. 

If we assume that $T^\dagger$ is small enough for all data points (which needs to be checked a posteriori), then the prediction of the double-frequency model, Eq. \ref{eq:impact2}, is precisely the same as the one of the standard LLOB model, up to a rescaling of the $x$-axis by $\eta^\star$, and of the $y$-axis by a ratio $\sqrt{D_\mathrm{s}J/DJ_\mathrm{s}}$. Fig. \ref{figDonier} shows that the LLOB scaling prediction indeed reproduces the data very well, which allows a direct determination of $\eta^\star \approx J_\mathrm{s}/J \approx 3.15 \times 10^{-3}$. In other words, we find that most of the daily liquidity is provided by ``fast'' agents, as expected. We have checked that the value of $\eta^\star$ is not significantly different in the period 2007-2008 and 2009-2010. We have also investigated the dependence of $\eta^\star$ on market capitalisation and volatility. We find that low volatility/large cap. stocks are characterized by a larger value of $\eta^\star$ than high volatility/small cap. stocks, suggesting, perhaps counter-intuitively, that the low frequency activity is comparatively more important in low volatility/large cap. stocks.  

The large $\eta$ plateau value, on the other hand, imposes $\sqrt{D_\mathrm{s}J/DJ_\mathrm{s}} = 0.4/\sqrt{2}$, leading to $\sqrt{D_\mathrm{s}/D} \simeq 10^{-2}$. Since $\sqrt{D}$ should be close to the price volatility \cite{Donier}, we find that, consistently with its interpretation, the ``slow'' liquidity moves much more slowly than the price itself. These estimates in turn lead to $T^\dagger \approx 45\, \nu_\mathrm{f}^{-1}$, or $\sim 45$ seconds for $\nu_\mathrm{f}^{-1}=1$ second. Since the median execution time of the metaorders in our sample is $35$ minutes, we conclude that most metaorders in our sample are indeed longer than $T^\dagger$. 

\begin{figure}[ht]
\includegraphics[width=0.8\linewidth]{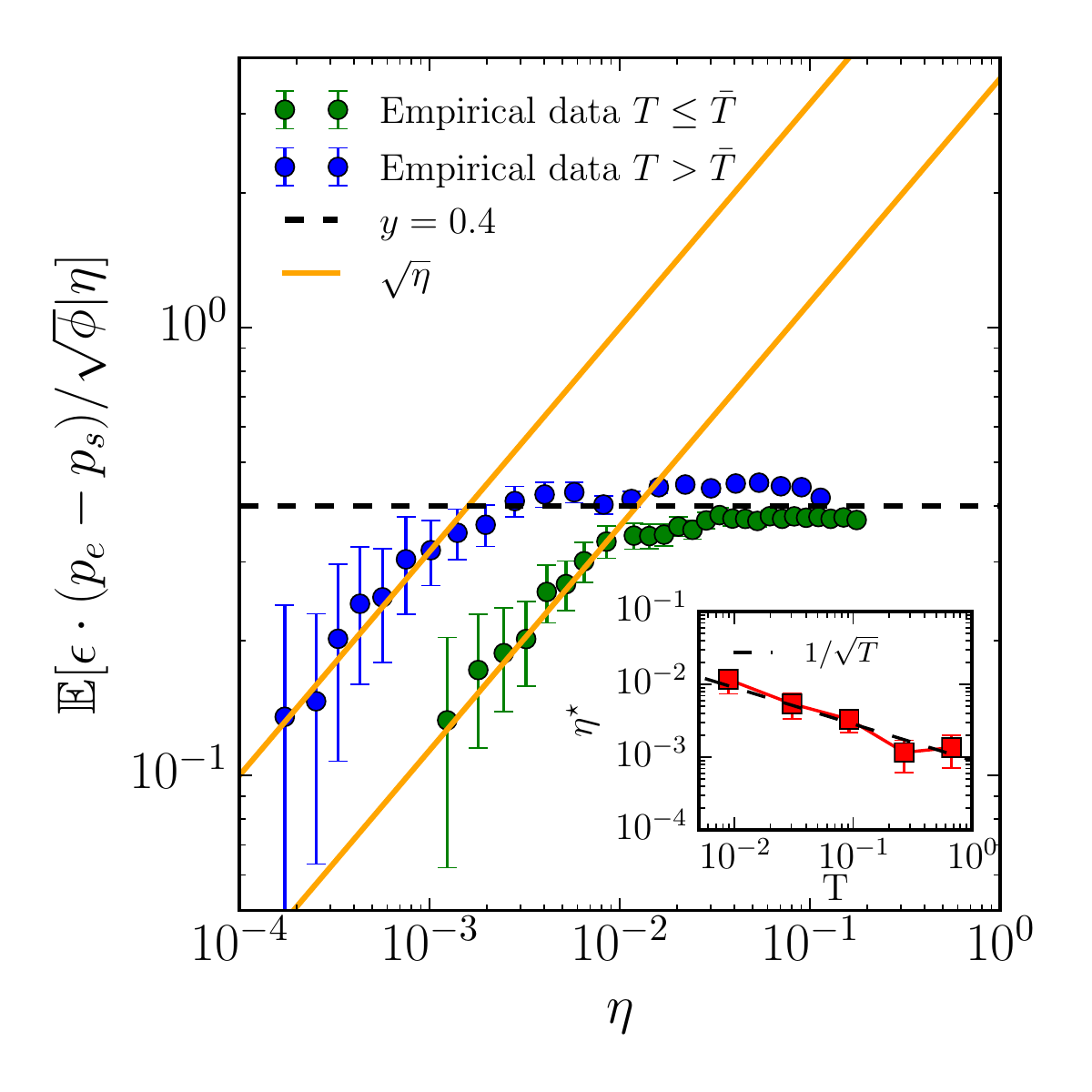}
\caption{(Main panel) Empirically determined scaling function ${\cal F}(\eta)$ vs. participation rate $\eta$ for metaorders with duration $T$ larger (blue dots) or smaller (green dots) than the median sample duration $\bar{T} \approx 0.09$. Note that the crossover value $\eta^\star$ is $\sim 10$ times larger in the latter case. Note that the asymptotic value of ${\cal F}$ (and hence the impact for a given $Q$) is approximately independent of $T$, as predicted by the theory. Both empirical curves are obtained using metaorders with sufficient large order size, i.e. $\phi \gtrsim 10^{-5}$. (Inset) Plot of the crossover participation rate $\eta^\star$ as a function of the execution time $T$, revealing an approximate $T^{-1/2}$ behaviour.}
\label{figCondD}
\end{figure}

Still, a bi-modal distribution of trading frequencies is certainly an oversimplification. One should consider instead, as in \cite{Benzaquen}, a continuous distribution of frequencies. Several empirical facts about the dynamics of financial markets (see e.g. \cite{LilloEPJB,Eisler,BacryHawkes,TQP}) actually suggest that such a distribution is a power-law. The numerical solution and the fitting procedure of such a general model is beyond the scope of the present paper, but the simplified analysis \cite{Benzaquen} suggests that the LLOB scaling function should be approximately valid, with a crossover value $\eta^\star$ that decreases as a power-law of $T$. Intuitively, the critical participation rate $\eta^{\star}$ should indeed be larger for small durations $T$, since there are less traders that can be considered ``fast'' on a such short time scales and more traders that are ``slow'' on the timescale of the metaorder. This intuition is indeed confirmed by Fig. \ref{figCondD}-top where we show the rescaled data as a function of $\eta$, for metaoders longer and shorter than the median execution time $\bar{T}\approx 0.09$. The crossover participation rate $\eta^{\star}$ for small durations is found to be 10 times larger for large durations. In Fig. \ref{figCondD}-bottom, we show the $T$-dependence of $\eta^\star$, obtained by fitting the rescaled data by $\mathcal{F}(\eta/\eta^\star)$ using five bins of $T$ containing the same number of data points ($\sim 1.4\times 10^6$), suggesting $\eta^\star \sim T^{-1/2}$ \footnote{Note that this behaviour is clearly distinct from the prediction of the two-frequency model, namely $\eta^\star \sim T^{-1}$ for $T < T^\dagger$ and $\sim$ constant for $T > T^\dagger$.}. It would be very interesting to use this result to map out the frequency distribution of the hidden liquidity, but this requires going beyond the approximate solution of \cite{Benzaquen}. We leave this for a subsequent investigation.  

In this paper, we have used a very large data set of orders executed in the US equity market to quantitatively test a market impact model, which predicts a crossover from a linear (in volume) behaviour for small volumes to a square-root behaviour for intermediate volumes. The data unambiguously suggests the existence of such a crossover, and once again confirms the square-root law which, as emphasized on several previous occasions, is remarkably independent of the execution time -- contradicting many early theories \cite{Torre,Zhang,Gabaix}. We have shown how the data points towards the existence of multiple time scales in the dynamics of liquidity, with its high frequency component dominating the total market activity and its low-frequency component contributing to the concavity of the impact function. (For an alternative and complementary viewpoint, see \cite{Fosset}). Our results are interesting from two rather different points of views. One is that they represent a significant improvement in our understanding of the determinants of market impact which is both the main component of trading costs for institutional investors and an important aspect of the stability of financial markets. The second aspect is that we are entering an era where economic and financial data becomes of such quality that theoretical ideas can be tested with standards comparable to those of natural sciences.  

We thank J. Donier for early discussions on this subject and Z. Eisler, J. Kockelkoren, C.-A. Lehalle, I. Mastromatteo, B. Toth and E. Zarinelli for very fruitful conversations.

\section*{Data availability statement}
The data were purchased by Imperial College from the company ANcerno Ltd (formerly the Abel Noser Corporation) which is a widely recognised consulting firm that works with institutional investors to monitor their equity trading costs. Its clients include many pension funds and asset managers. The authors do not have permission to redistribute them, even in aggregate form. Requests for this commercial dataset can be addressed directly to the data vendor.  See {\tt www.ancerno.com} for details.

% Create the reference section using BibTeX:
%\bibliography{basename of .bib file}

\begin{thebibliography}{4}
\expandafter\ifx\csname natexlab\endcsname\relax\def\natexlab#1{#1}\fi
\expandafter\ifx\csname bibnamefont\endcsname\relax
  \def\bibnamefont#1{#1}\fi
\expandafter\ifx\csname bibfnamefont\endcsname\relax
  \def\bibfnamefont#1{#1}\fi
\expandafter\ifx\csname citenamefont\endcsname\relax
  \def\citenamefont#1{#1}\fi
\expandafter\ifx\csname url\endcsname\relax
  \def\url#1{\texttt{#1}}\fi
\expandafter\ifx\csname urlprefix\endcsname\relax\def\urlprefix{URL }\fi
\providecommand{\bibinfo}[2]{#2}
\providecommand{\eprint}[2][]{\url{#2}}

\bibitem[{Note1()}]{Note1}
Note1, \bibinfo{note}{\uppercase {A}Ncerno Ltd (formerly the Abel Noser
  Corporation) is a widely recognized consulting firm that works with
  institutional investors to monitor their equity trading costs. Its clients
  include many pension funds and assets managers. Previous academic studies
  that use ANcerno data to investigate the market impact at different times
  scales includes \cite {Zarinelli, Bucci}. See {\protect \tt www.ancerno.com}
  for details.}

\bibitem[{Note2()}]{Note2}
Note2, \bibinfo{note}{the sample represents around the 5\% of the total market
  volume}.

\bibitem[{Note3()}]{Note3}
Note3, \bibinfo{note}{directly regressing the impact as $\protect \sqrt {\phi }
  T^{-\beta }$ in the $\eta > \eta ^\star $ regime yields $\beta = - 0.04 \pm
  0.02$, confirming the near independence of the impact on the execution time
  $T$; {while in the $\eta < \eta ^\star $ regime the regression as $\phi
  T^{-\beta }$ yields $\beta = 0.45 \pm 0.05$, in close agreement with the LLOB
  prediction $\beta =1/2$.}}

\bibitem[{Note4()}]{Note4}
Note4, \bibinfo{note}{note that this behaviour is clearly distinct from the
  prediction of the two-frequency model, namely $\eta ^\star \sim T^{-1}$ for
  $T < T^\dagger $ and $\sim $ constant for $T > T^\dagger $.}

\end{thebibliography}


\begin{thebibliography}{100}
\bibitem{Takayazu} K. Kanazawa, T. Sueshige, H. Takayasu, \& M. Takayasu, \textit{Derivation of the Boltzmann equation for financial Brownian motion: Direct observation of the collective motion of high-frequency traders}. Physical review letters, 120(13), 138301 (2018).
\bibitem{Kyle}A. S. Kyle, \textit{Continuous auctions and insider trading.} Econometrica, pp.
1315-1335, (1985).
\bibitem{Torre}N. Torre, \textit{Barra market impact model handbook.} BARRA Inc., Berkeley, 1997.
\bibitem{Almgren}R. Almgren, C. Thum, E. Hauptmann, \& H. Li, \textit{Direct estimation of equity market impact.} Risk, vol. 57, (2005).
\bibitem{Engle}R. F. Engle, R. Ferstenberg, \& J. Russell, \textit{Measuring and modeling execution cost and risk.} Chicago GSB Research Paper, no. 08-09, (2008).
\bibitem{Moro} {E. Moro, J. Vicente, L.G. Moyano, A. Gerig, J.D. Farmer,
G. Vaglica, F. Lillo, \& R.N. Mantegna}, \textit{Market impact and trading profile of hidden orders in stock markets}. Phys. Rev. E  80, 066102 (2009).
\bibitem{Toth}B. T\'oth, Y. Lemperiere, C. Deremble, J. De Lataillade, J. Kockelkoren, \&
J.-P. Bouchaud, \textit{Anomalous price impact and the critical nature of liquidity in financial markets.} Physical Review X, vol. 1, no. 2, p. 021006, (2011).
\bibitem{Brokmann} Brokmann, X., Serie, E., Kockelkoren, J., \& Bouchaud, J. P. \textit{Slow decay of impact in equity markets}. Market Microstructure and Liquidity, 1(02), 1550007 (2015).
\bibitem{Zarinelli} E. Zarinelli, M. Treccani, J. D. Farmer \& F. Lillo, \textit{Beyond the Square Root: Evidence for Logarithmic Dependence of Market Impact on Size and Participation Rate}. Market Microstructure and Liquidity, Vol. 1, No. 2, (2015).
\bibitem{Bacry}E. Bacry, A. Iuga, M. Lasnier, \& C. A. Lehalle, C. A., \textit{Market impacts and the life cycle of investors orders}. Market Microstructure and Liquidity, 1(02), 1550009, (2015).
\bibitem{Bonart} J. Donier, \& J. Bonart, \textit{ A Million Metaorder Analysis of Market Impact on the Bitcoin}, Market Microstructure and Liquidity, Vol. 01, No. 02, (2015).
\bibitem{Toth2}B. Toth, Z. Eisler, \& J.-P. Bouchaud, \textit{The square-root impact law also holds for option markets}, Wilmott (2016).
\bibitem{Bucci} F. Bucci, I. Mastromatteo, Z. Eisler, F. Lillo, J.-P. Bouchaud, \& C.-A. Lehalle, \textit{Co-impact: Crowding effects in institutional trading activity}, https://arxiv.org/abs/1804.09565, (2018).
\bibitem{Frazzini} A. Frazzini, R. Israel, \& T. J. Moskowitz, \textit{Trading Costs}, https://ssrn.com/abstract=3229719, (2018).
\bibitem{TQP}J.-P. Bouchaud, J. Bonart, J. Donier, \& M. Gould, \textit{Trades, Quotes and Prices: Financial Markets Under The Microscope} Cambridge University Press, (2018). 
\bibitem{MTB}I. Mastromatteo, B. T\'oth, \& J.-P. Bouchaud, \textit{Anomalous impact in reaction-diffusion financial models} Physical Review Letters, 113(26), 268701, (2014).
\bibitem{Donier} J. Donier, J. Bonart, I. Mastromatteo, \& J.-P. Bouchaud, \textit{A fully consistent, minimal model for non-linear market impact}, Quantitative Finance 15, 1009-1121, (2015).
\bibitem{Benzaquen} M. Benzaquen, \& J.-P. Bouchaud, \textit{Market impact with multi-timescale liquidity}, Quantitative Finance, 2018, p. 1-10, (2018).
\bibitem{CFMunpub} CFM execution team, unpublished reports.
\bibitem{Zhang} Y. C. Zhang, \textit{Toward a theory of marginally efficient markets.} Physica A: Statistical Mechanics and its Applications, 269(1), 30-44 (1999).
\bibitem{Gabaix} X. Gabaix, P. Gopikrishnan, V. Plerou, \& H. E. Stanley, \textit{A theory of power-law distributions in financial market fluctuations.} Nature, 423(6937), 267-270, (2003).
\bibitem{LilloEPJB} F. Lillo, \textit{Limit order placement as an utility maximization problem and the origin of power law distribution of limit order prices}. {European Physical Journal B}  55, 453 (2007).
\bibitem{Eisler} Z. Eisler, J. Kertesz, F. Lillo \& R.N. Mantegna, \textit{Diffusive behavior and the modeling of characteristic times in limit order executions}. {Quantitative Finance} 9 547 (2009).
\bibitem{BacryHawkes} E. Bacry, K. Dayri, \& J. F. Muzy, \textit{Non-parametric kernel estimation
for symmetric Hawkes processes}. Application to high frequency financial data. The European Physical Journal B-Condensed Matter and Complex Systems, 85(5), 1-12 (2012).
\bibitem{Fosset} L. Dall'Amico, A. Fosset, J.-P. Bouchaud, \& M. Benzaquen, \textit{How does latent liquidity get revealed in the limit order book?} arXiv 1808.09677, (2018).


\end{thebibliography}
%\bibliographystyle{h-physrev}

\end{document}